    \documentclass[prd,aps,preprint,tightenlines,showpacs,nofootinbib,superscriptaddress]{revtex4-1}
\usepackage{mathrsfs}
\usepackage{amsfonts}
\usepackage{amsmath}
\usepackage{amssymb}
\usepackage{array}
\usepackage{verbatim}
\usepackage{bm}
\usepackage{epsfig}
\usepackage{graphicx,color}
\usepackage{relsize}
\usepackage{lineno}
\usepackage{float}
\usepackage{multirow}
\RequirePackage{xspace}

\def\lsim{\raise0.3ex\hbox{$<$\kern-0.75em\raise-1.1ex\hbox{$\sim$}}}

\def\gsim{\raise0.3ex\hbox{$>$\kern-0.75em\raise-1.1ex\hbox{$\sim$}}}

\newcommand{\be}{\begin{equation}}

\newcommand{\ee}{\end{equation}}

\def\beq{\begin{equation}}

\def\eeq{\end{equation}}

\def\beqa{\begin{eqnarray}}

\def\eeqa{\end{eqnarray}}

\newcommand{\ba}{\begin{eqnarray}}

\newcommand{\ea}{\end{eqnarray}}

\def\gappeq{\mathrel{\rlap {\raise.5ex\hbox{$>$}}

{\lower.5ex\hbox{$\sim$}}}}

\def\lappeq{\mathrel{\rlap{\raise.5ex\hbox{$<$}}

{\lower.5ex\hbox{$\sim$}}}}

\def\Toprel#1\over#2{\mathrel{\mathop{#2}\limits^{#1}}}

\begin{document}

\title{Photoproduction of QED bound states in future \\ electron-ion colliders}

\author{Reinaldo {\sc Francener}}
\email{reinaldofrancener@gmail.com}
\affiliation{Instituto de Física Gleb Wataghin - UNICAMP, 13083-859, Campinas, SP, Brazil. }

\author{Victor P. {\sc Gon\c{c}alves}}
\email{barros@ufpel.edu.br}
\affiliation{Institute of Physics and Mathematics, Federal University of Pelotas, \\
  Postal Code 354,  96010-900, Pelotas, RS, Brazil}
\affiliation{Institute of Modern Physics, Chinese Academy of Sciences,
  Lanzhou 730000, China}

\author{Bruno D. {\sc Moreira}}
\email{bruno.moreira@udesc.br}
\affiliation{Departamento de F\'isica, Universidade do Estado de Santa Catarina, 89219-710 Joinville, SC, Brazil. }

\author{Kaique A. {\sc Santos}}
\email{kaique.santos@edu.udesc.br}
\affiliation{Departamento de F\'isica, Universidade do Estado de Santa Catarina, 89219-710 Joinville, SC, Brazil. }

\begin{abstract}
 In this work we perform an exploratory study of the photoproduction of singlet QED bound states $(l^-l^+)_S$ in electron - ion collisions at the EicC, EIC and LHeC energies. The total cross - sections, event rates per year and rapidity distributions associated with the parapositronium, paramuonium and paratauonium production are estimated.  Moreover, we  consider the decay of these states in a two - photon system and implement kinematical cuts on the rapidities and energies of the photons in the final state. We demonstrate the paramuonium can, in principle, be observed for the first time in all these colliders and that the EIC is a potential collider to discover the paratauonium state.
\end{abstract}



\maketitle

\vspace{1cm}

\begin{figure}[b]
	\centering
\includegraphics[width=0.65\textwidth]{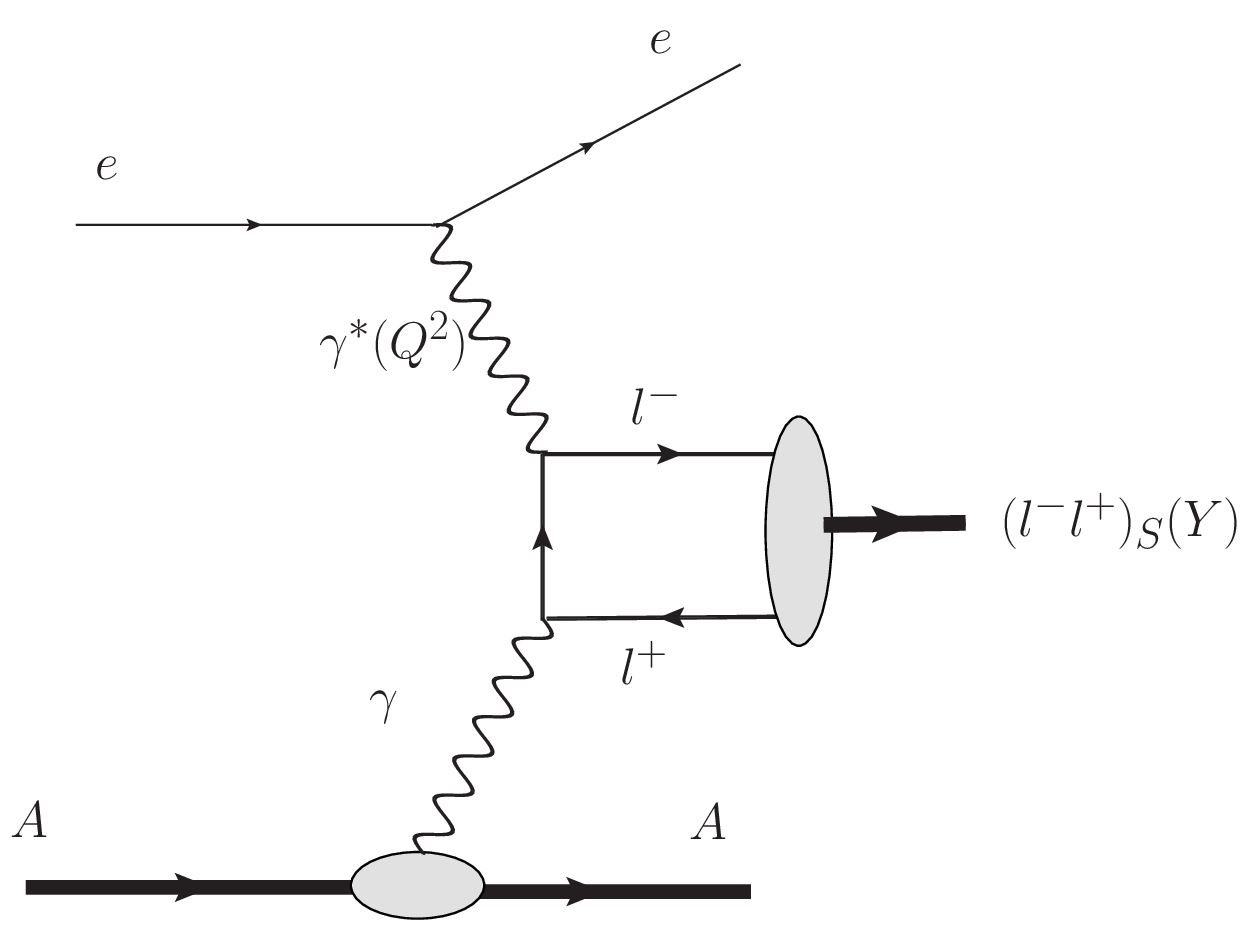} 
\caption{Production of a singlet QED bound state by $\gamma^* \gamma$ interactions in eA  collisions. }
\label{fig:diagram}
\end{figure}

%\section{Introduction}
Electron - hadron colliders are the ideal facilities to improve our understanding of the strong interactions
theory – the Quantum Chromodynamics (QCD). In particular, the study of production and propagation of QCD bound states in a nuclear medium is  one of the major motivations for the construction of the Electron - Ion Collider (EIC) in the USA \cite{eic}, recently approved, as well as for the proposal of future electron – hadron colliders at CERN \cite{lhec} and China \cite{EicC}. Due to the large luminosities expected to be present in  these colliders, they will allow the investigation of the hadronic structure with unprecedented precision
to inclusive and diffractive observables. More recently, such high luminosities have also motivated the searching for beyond the Standard Model (BSM) Physics in these colliders (See, e.g. Refs. \cite{Azuelos:2019bwg,Boughezal:2022pmb,Cottin:2021tfo,Davoudiasl:2021mjy,Liu:2021lan,Batell:2022ogj,Gu:2022nlj,Balkin:2023gya}). In this letter, we will investigate, for the first time, the possibility to study the production of QED bound states by two - photon fusion in $eA$ colliders, as represented in Fig. \ref{fig:diagram}.  Such a physical system is characterized by a bound state formed by leptons of a same flavor, denoted hereafter by $(l^+l^-)$, which can be in a singlet state with spins antiparallel and total spin $s=0$, or it can be in the triplet state with spins parallel and total spin $s=1$. Such singlet ($S$) and triplet ($T$) states are usually denoted para and ortho QED bound states, respectively (For a recent review, see Ref. \cite{Adkins:2022omi}). We will focus on para QED bound states, which can be produced at the Born level by two - photon fusion. Over the last years, the production of QED bound states in $e^+ e^-$ and hadronic colliders was investigated by several authors \cite{serbo_jtep,serbo_pra,serbo_muonium,serbo_posi,nos_muonium,Francener:2021wzx,dEnterria:2022ysg,dEnterria:2023yao,Dai:2024imb,Bertulani:2023nch}, strongly motivated by the possibility of probing the  paramuonium $(\mu^- \mu^+)_S$ and  paratauonium $(\tau^- \tau^+)_S$ states, which have never been observed \footnote{Such states are also denoted true muonium and true tauonium in the literature \cite{nos_muonium,Francener:2021wzx,dEnterria:2022ysg,dEnterria:2023yao}.}. In particular, in Ref. \cite{Francener:2021wzx}, the production of these states in ultraperipheral heavy ion collisions at RHIC, LHC and FCC energies was discussed in detail. 
Our goal in this letter is to complement these previous studies and estimate the total cross-sections and event rates associated with  the photoproduction of para QED bound states in $eA$ collisions at the EIC, LHeC and EicC energies and luminosities. As we will demonstrate in what follows, our results derived including realistic experimental cuts indicate that a future analysis of  $(\mu^- \mu^+)_S$ and $(\tau^- \tau^+)_S$ states is, in principle, feasible in these future colliders.

Initially, let's present a brief review of the formalism needed to estimate the photoproduction of QED bound states in $eA$ colliders. Assuming the validity of the equivalent photon approximation \cite{epa}, the associated total cross-section for the photoproduction of a singlet (S)  QED bound states can be expressed by
\begin{eqnarray}
\sigma \left[e A \rightarrow e \otimes (l^+ l^-)_S \otimes  A;s_{NN} \right]   
&=& \int \mbox{d}\omega_e 
\mbox{d}\omega_A \,   f_{\gamma/e}(\omega_e) f_{\gamma/A}(\omega_A)  \, \hat{\sigma}\left[\gamma \gamma \rightarrow (l^+ l^-)_S ; 
W_{\gamma \gamma} \right] 
  \,\,\, ,
\label{Eq:cs_singlet}
\end{eqnarray}
where $\otimes$ represents a rapidity gap in the final state, $\sqrt{s}$ is center - of - mass energy of the $eA$ collision, $f_{\gamma/i}$ is the distribution function  associated with  the flux of photons generated by the particle $i$ ($i = e, \,A$) with an energy  $\omega_i$. Moreover,   $\hat{\sigma}$ represents the cross-section for the photon - induced interaction for a  given photon - photon center - of - mass energy $W_{\gamma \gamma}$. At the Born level, the cross-section $\hat{\sigma}_{\gamma \gamma}$ for the photoproduction of a singlet QED bound state in $\gamma \gamma$ interactions can be estimated using the Low formula \cite{Low}, which allow us to express this cross-section in terms of the two-photon decay width $\Gamma_{(l^+ l^-)_S \rightarrow \gamma \gamma}$ as  follows
\begin{eqnarray}
 \hat{\sigma}_{\gamma \gamma \rightarrow (l^+ l^-)_S}(\omega_{e},\omega_{A}) = 
8\pi^{2} (2J+1) \frac{\Gamma_{(l^+ l^-)_S \rightarrow \gamma \gamma}}{M} 
\delta(4\omega_{e}\omega_{A} - M^{2}) \, ,
\label{Low_cs}
\end{eqnarray}
where $M = 2 m_{l}$ and $J$ are, respectively, the mass and spin of the  produced singlet QED state. In the non -- relativistic approximation, one have that only the probability density of $s$ -- states at the origin does not vanish, which implies that $|\Psi_{ns}(0)|^2 = \alpha^3 m_{l}^3/8 \pi n^3$ \cite{Adkins:2022omi}, where $\alpha$ is the electromagnetic coupling constant. Consequently, one has $\Gamma (n \,^1S_0) = \alpha^5 m_{l}/2 n^3$ and that the $\gamma \gamma$ cross-section for the lowest singlet QED bound state is given by
\begin{eqnarray}
 \hat{\sigma}_{\gamma \gamma \rightarrow (l^+ l^-)_S}(\omega_{e},\omega_{A}) = 2 \pi^2 \alpha^5 \delta(4\omega_{e}\omega_{A} - M^{2})\,\,.
\end{eqnarray}
For the photon distribution functions, we will assume that the flux associated with the electron is given by \cite{epa}
\begin{eqnarray}
    f_{\gamma/e}(\omega_e) =    \frac{\alpha_{em}}{\pi \omega_e} \int \frac{\mbox{d} Q^2}{Q^2}   \left[\left(1 - \frac{\omega_e}{E_e}\right)\left(1 - \frac{Q^2_{min}}{Q^2}\right) + \frac{\omega_e^2}{2E_e^2}\right]\,,
\end{eqnarray}
where $\omega_e$ is the energy of the photon emitted by the electron with energy $E_e$, and $Q^2$ is its virtuality. Moreover, one has that $Q_{min}^2 = m_e^2\omega_e^2/[E_e(E_e - \omega_e)]$ and $Q_{max}^2 = 4E_e(E_e - \omega_e)$, which is constrained by the maximum of the electron energy loss. As we are interested in the photoproduction of QED bound states, we will assume $Q_{max}^2 = 1.0$ GeV$^2$.
On the other hand, the equivalent photon distribution for a nucleus will be estimated as follows,
\begin{eqnarray}
    f_{\gamma/A}(\omega_A) = \frac{Z^{2}\alpha}{\pi^2}  \int \mbox{d}^2r \frac{1}{r^{2} v^{2}\omega_A}
 \cdot \left[
\int u^{2} J_{1}(u) 
F\left(
 \sqrt{\frac{\left( \frac{r\omega_A}{\gamma_L}\right)^{2} + u^{2}}{r^{2}}}
 \right )
\frac{1}{\left(\frac{r\omega_A}{\gamma_L}\right)^{2} + u^{2}} \mbox{d}u
\right]^{2} \,\,,
\label{fluxo}
\end{eqnarray}
where $F(q)$ is the  charge form factor,   $\gamma_L$ is the Lorentz factor and $v$ is the nucleus velocity. We will estimate the nuclear photon spectrum considering the realistic form factor, which corresponds to the Wood - Saxon distribution and is the Fourier transform of the charge density of the nucleus, constrained by the experimental data.

%\section{Results}

In what follows, we will estimate the total cross-sections and rapidity considering the energy and target configurations expected in the future electron-ion colliders at the BNL, CERN and in China. As the future electron-ion collider (EIC) at BNL,  the electron beam with an energy up to 18 GeV will be set to collide  with a heavy ion with energies up to 100 GeV \cite{eic}, reaching  integrated luminosities in the $10 - 100$ fb$^{-1}$ range. In our analysis for the EIC, we will assume  the following  benchmarks:   $(E_e,\, E_{Au}) = (18,\, 100)$ GeV and ${\cal{L}} = 100$ fb$^{-1}$.  Moreover, we will also estimate the distributions for the EicC \cite{EicC} ($E_e = 3.5$ GeV, $E_{Au} = 10$ GeV and ${\cal{L}} = 50$ fb$^{-1})$ and for the  LHeC \cite{lhec} ($E_e = 50$ GeV, $E_{Pb} = 2760$ GeV and   ${\cal{L}} = 1.0$ fb$^{-1}$).

\begin{figure}[t]
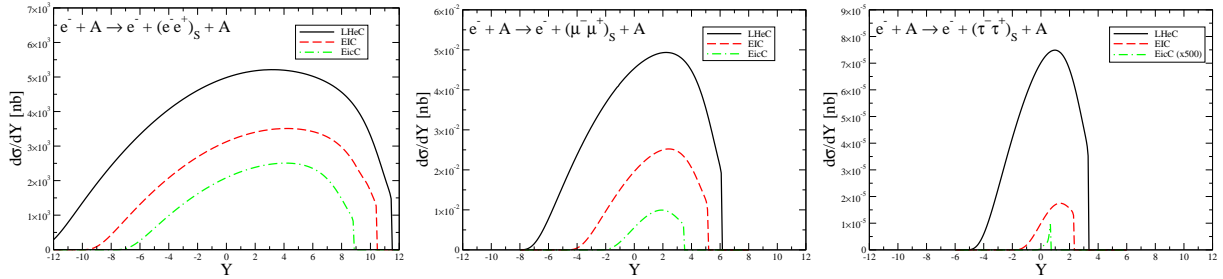

	\centering
	\begin{tabular}{ccccc}
\includegraphics[width=0.32\textwidth]{dsdY_positronium_eA_nocuts.eps} &
\includegraphics[width=0.32\textwidth]{dsdY_muonium_eA_nocuts.eps} &
\includegraphics[width=0.32\textwidth]{dsdY_tauonium_eA_nocuts.eps} 
	\end{tabular}
\caption{ Rapidity distributions for the photoproduction of singlet QED bound states in eA  collisions. Predictions for the parapositronium (left), paramuonium (central) and paratauonium (right) production at the EIC, LHeC and EicC energies. }
\label{fig:dsdY}
\end{figure}

\begin{center}
	\begin{table}[t]
		\begin{tabular}{|c|c|c|c|c|c|c|}
			\hline
			\hline 
			& {\bf Parapositronium}          &       {\bf Paramuonium} &    {\bf Paratauonium}   \\
			\hline 	
			\hline 
eAu (EicC)  & 27.06$\times 10^{3}$ (1.15$\times 10^{12}$) & 33.96$\times 10^{-3}$ (1.70$\times 10^{6}$) & 4.58$\times 10^{-9}$ (0.23) \tabularnewline
			\hline 
eAu (EIC)& 48.45$\times 10^{3}$ (4.84$\times 10^{12}$) & 150.88$\times 10^{-3}$ (15.09$\times 10^{6}$) & 41.44$\times 10^{-6}$   (4.14$\times 10^{3}$) \tabularnewline
			\hline 
ePb (LHeC) & 89.13$\times 10^{3}$ (89.13$\times 10^{9}$) & 443.41$\times 10^{-3}$ (443.41$\times 10^{3}$) & 380.41$\times 10^{-6}$ (380.41) \tabularnewline
			\hline 
			\hline 			
		\end{tabular}
		\caption{Total cross-sections in nb (Event rates per year)  for the production of singlet QED bound states in eA collisions.}
		\label{table:total_nocuts}
	\end{table}
\end{center}

In Fig. \ref{fig:dsdY} we present our predictions for the rapidity distributions associated with the photoproduction of  parapositronium (left), paramuonium (central) and paratauonium (right)  at the EicC, EIC and LHeC energies. Such distributions can be derived using that in the $eA$ cm-frame, the photon energies $\omega_i$ can be written in terms of the rapidity
$Y$ of the QED bound state as follows 
\begin{eqnarray}
    \omega_e = \frac{M}{2} e^{+Y}   \,\,\,\,\,\, \mbox{and} \,\,\,\,\,\,  \omega_A = \frac{M}{2} e^{-Y} \,.
    \label{eq:omegas}
\end{eqnarray}
One has that  the distribution increases with energy, with the maximum occurring at larger values of $Y$ when the center-of-mass energies are increased. Moreover, it is wider and have a larger normalization for lighter QED bound states. The associated predictions for the total cross - sections and event rates per year are presented in Table \ref{table:total_nocuts}. For the parapositronium, we predict cross - sections of the order of $\mu$b and $\approx 10^{12}\, (10^{9}\,)$ events per year at the EIC and EicC (LHeC). On the other hand, for the paramuonium (paratauonium), the predictions are reduced by a factor $10^6$ ($10^9$), except the results for the paratauonium production at the EicC, where the suppression is larger due to the small phase space available for the production of this massive QED bound state. Such results strongly motivate the analysis of the experimental separation of the events.

\begin{center}
	\begin{table}[t]
		\begin{tabular}{|c|c|c|c|c|}
			\hline
			\hline 
			&    \multicolumn{2}{c|}{\bf  $ eA \rightarrow e(\mu^- \mu^+)_SA \rightarrow e \gamma \gamma A$ } &        \multicolumn{2}{c|}{\bf $ eA \rightarrow e(\tau^- \tau^+)_S A \rightarrow e \gamma \gamma A$ }   \\
			\hline 	
			\hline 
      &       $E_{\gamma_{1}},E_{\gamma_{2}}\ge 1.0\,$GeV  & $E_{\gamma_{1}},E_{\gamma_{2}} \ge 0.5\,$GeV &    $E_{\gamma_{1}},E_{\gamma_{2}} \ge 1.0\,$GeV  & $E_{\gamma_{1}},E_{\gamma_{2}} \ge 0.50 $ 
   \\
			\hline 	
eAu (EicC)  &  0.50$\times 10^{-3}$ (0.025$\times 10^{6}$) & 2.98$\times 10^{-3}$ (0.15$\times 10^{6}$) & 3.05$\times 10^{-9}$ (0.15) & 3.09$\times 10^{-9}$ (0.15) 
\tabularnewline
			\hline 
eAu (EIC)&  3.18$\times 10^{-3}$ (0.32$\times 10^{6}$) &  12.63$\times 10^{-3}$ (1.26$\times 10^{6}$)  & 28.80$\times 10^{-6}$   (2.88$\times 10^{3}$) & 31.02$\times 10^{-6}$ (3.10$\times 10^{3}$)
\tabularnewline
			\hline 
ePb (LHeC) & 9.33$\times 10^{-3}$ (9.33$\times 10^{3}$) & 34.55$\times 10^{-3}$ (34.55$\times 10^{3}$) & 240.29$\times 10^{-6}$ (240.29) & 256.07$\times 10^{-6}$ (256.07) 
\tabularnewline
			\hline 
			\hline 			
		\end{tabular}
		\caption{Cross-sections in nb (Event rates per year) for the production of a two - photon system associated with the decay of a singlet QED bound state created in eA collisions, derived considering cuts in pseudorapidity and energy of each photon after decay.}
		\label{table:withcuts}
	\end{table}
\end{center}

After being produced, the QED bound states mainly decay back in photons. As a consequence, 
the final state will be characterized by the presence of an electron, the ion, the two - photon system and   two rapidity gaps, i.e. empty regions  in pseudorapidity that separate the intact very forward electron and nucleus from the $\gamma \gamma$ system. Moreover, the invariant mass of the two - photon system will have a peak for $m_{\gamma \gamma} \approx M$, which strongly enhance the signal with relation to potential backgrounds, as e.g. the continuum associated with the Light - by - Light scattering (See, e.g. Ref. \cite{Balkin:2023gya}   for a recent discussion about this topic). Such aspects can, in principle, be used to separate the events and to probe the QED bound states. In what follows, we will analyze the impact of kinematical cuts in the photon pseudorapidities and energies  on our predictions. In particular, we will consider the following cuts:
\begin{eqnarray}
    |\eta_1|, |\eta_2|  \le  3.5  \,\,\,\,\,\, \mbox{and} \,\,\,\,\,\,
    E_{\gamma_1}, E_{\gamma_2}  \ge  1.0 \,\, \mbox{GeV}   \,\,,
\end{eqnarray}
which are expected to be valid at the EIC and are associated with the acceptance region determined by the EIC calorimeters  \cite{eic}. These two cuts ensure that the photons can be detected by the central detector and  are energetic enough to be reconstructed in the calorimeter. These cuts will also be applied in the predictions for the EicC and LHeC. In addition, we will investigate how sensitive are our predictions to the minimum value for the photon energy, and also present results for $E_{\gamma_1}, E_{\gamma_2}  >  0.5 \,\, \mbox{GeV}$. The decay of the QED bound state will be simulated using the Pythia6 event generator \cite{Sjostrand:2006za}. It is important to emphasize that these cuts are very restrictive for the photoproduction of parapositroniums, removing all signal, which is directly associated with the very small mass of this state. An alternative to probe the parapositronium in electron - ion collisions is to consider its electroproduction, characterized by $Q^2 \gg 1.0$ GeV$^2$, in which the $(e^-e^+)_S$ state is produced with a large transverse momentum. Such possibility will be investigated in a separated publication.
In Table \ref{table:withcuts} we present the resulting predictions for the total cross - sections and events rates per year associated with the production of a two - photon system  in the decay of a paramuonium or paratauonium created in a $eA$ collision. We present the results for two values of the minimum photon energy. One has that the reduction of $E_{\gamma}^{min}$ increases the  paramuonium signal by a factor 4. In contrast, such a cut has a small impact on the prediction for the paratauonium.
In comparison with the results presented in Table \ref{table:total_nocuts}, one has that the cuts and the decay imply the suppression in the number of events associated with the paramuonium (paratruonium) by a factor $\ge 10$ (1.3). However, the results presented in Table \ref{table:withcuts} indicate that the analysis of the paramuonium is, in principle, feasible in $eA$ collisions at the EicC, EIC and LHeC, allowing to observe this final state for the first time. On the other hand, for the paratauonium case, the EIC is the potential collider to discover this final state and searching e.g. for signals of BSM physics, which are expected to have a larger impact on heavier leptons.

As a summary, in this letter we have performed an exploratory study of the photoproduction of QED bound states in electron - ion collisions at the EicC, EIC and LHeC energies. We have estimated the total cross - sections, event rates per year and rapidity distributions and demonstrated that a large number of events are expected in the future colliders. Moreover, we have considered the decay of these states in a two - photon system and implemented kinematical cuts on the rapidities and energies of the photons in the final state. We have demonstrated the paramuonium can, in principle, be observed for the first time in these colliders and that the EIC is a potential collider to discover the paratauonium state. The results derived here strongly motivate the implementation of this process in a Monte Carlo dedicated to the description of $eA$ collisions, as well as the extension of the study for the electroproduction of the QED bound states. Both subjects will be explored in forthcoming publications.

%\section{Summary}

\begin{acknowledgments}

R. F. acknowledges support from the Conselho Nacional de Desenvolvimento Científico e Tecnológico (CNPq, Brazil), Grant No. 161770/2022-3. V.P.G. was partially supported by CNPq, FAPERGS and INCT-FNA (Process No. 464898/2014-5).  B. D. Moreira and K. A. Santos were partially supported by FAPESC. 

\end{acknowledgments}

\hspace{1.0cm}

\end{document}